\documentclass[aps,showpacs,preprintnumbers,amsmath,amssymb,prl,twocolumn,superscriptaddress]{revtex4-1}
\usepackage{txfonts}
\usepackage{graphicx}
\usepackage{dcolumn}
\usepackage{bm}
\usepackage{color}
\def\comment#1{}

\def\slashchar#1{\setbox0=\hbox{$#1$}           
   \dimen0=\wd0                                 
   \setbox1=\hbox{/} \dimen1=\wd1               
   \ifdim\dimen0>\dimen1                        
      \rlap{\hbox to \dimen0{\hfil/\hfil}}      
      #1                                        
   \else                                        
      \rlap{\hbox to \dimen1{\hfil$#1$\hfil}}   
      /                                         
   \fi}                                         %

\def\sigmab{{\mbox{\boldmath $\sigma$}}}

\begin{document}
 
\title{Deconfined quantum criticality and logarithmic violations of scaling from emergent gauge symmetry}

\author{Flavio S. Nogueira}
\affiliation{Institut f{\"u}r Theoretische Physik III, Ruhr-Universit\"at Bochum,
Universit\"atsstra\ss e 150, 44801 Bochum, Germany}

\author{Asle Sudb{\o}}
\affiliation{Department of Physics, Norwegian University of
Science and Technology, N-7491 Trondheim, Norway}

\date{Received \today}

\begin{abstract}
We demonstrate that the low-energy effective theory for a deconfined quantum critical point in $d=2+1$ dimensions contains a leading 
order contribution given by the Faddeev-Skyrme model. The Faddeev-Skyrme term is shown to give rise to the crucial Maxwell term in the 
CP$^1$ field theory governing the deconfined quantum critical point. We derive the leading contribution to the spin stiffness near the 
quantum critical point and show that it exhibits a logarithmic correction to scaling of the same type as recently observed numerically 
in low dimensional models of quantum spin systems featuring a quantum critical point separating an antiferromagnetically ordered state 
from a valence bond solid state. These corrections, appearing away from upper or lower critical dimensions, reflect an emergent gauge 
symmetry of low-dimensional antiferromagnetic quantum spin systems. 
\end{abstract}

\maketitle

Much of our understanding of phase transitions is based on the concept of 
spontaneous symmetry breaking \cite{Coleman}, which provides a mechanism for the spontaneous 
generation of an ordered state as one or more parameters of a many-body system are varied. 
In Abelian systems the ordered state can be related to the disordered symmetric state by a duality 
transformation mapping a strongly coupled regime onto a weakly coupled one \cite{Savit,Kleinert-GFCM-1,Sudbo-Book}. 
In the case of a $U(1)$ symmetry, the symmetric phase is described in the dual picture by a {\it disorder} parameter 
\cite{Kleinert-GFCM-1}, as opposed to the order parameter describing the broken symmetry state in the original picture. 
The disorder parameter is nonzero when topological defects of the $U(1)$-theory (vortices) are condensed. The $U(1)$ 
symmetry of the dual theory is then spontaneously broken. The superfluid phase corresponds to the $U(1)$ symmetric 
state. The vortex condensation of the dual theory reflects the nontriviality of the first homotopy group of 
$U(1)$, namely  $\pi_1(U(1))=\mathbb{Z}$, the group of integers. This leads, for instance, to flux 
quantization in superconductors. 

When the order field is composed of more elementary constituents, the disordered phase exhibits non-trivial 
features which do not follow from standard spontaneous symmetry breaking arguments. This is so, for instance, 
in two-dimensional quantum spin systems featuring a paramagnetic phase where the symmetries of the underlying 
lattice is broken. On a square lattice where $SU(2)$-invariant spin interactions compete, a valence-bond solid 
(VBS) state emerges in the paramagnetic phase \cite{Sachdev-Review}. An example of this is the $J-Q$ model 
\cite{Sandvik_2007}, with a four-spin exchange around the plaquette of a square lattice in addition to the usual 
Heisenberg term,  
\begin{equation}
\label{J-Q}
 H=J\sum_{\langle i,j\rangle}{\bf S}_i\cdot{\bf S}_j -Q\sum_{\langle ijkl\rangle}\left({\bf S}_i\cdot{\bf S}_j-\frac{1}{4}\right)
\left({\bf S}_k\cdot{\bf S}_l-\frac{1}{4}\right),
\end{equation}
When $J\gg Q$ the Heisenberg term dominates the physics and the ground state is antiferromagnetic (AF), while for 
$Q\gg J$ the four-spin term favors a VBS state. Numerical works \cite{Sandvik_2007,Melko-Kaul_2008,Jiang_2008,Sandvik-2010} 
show that the $J-Q$ model has  an emergent $U(1)$ symmetry. This had previously been predicted in models with the same 
phase structure by introducing a new paradigm for phase transitions \cite{Senthil-2004}, the so called deconfined quantum criticality 
(DQC) scenario. Introducing two different order parameters to describe a phase transition, one for the AF phase and another one for the 
VBS phase, it was argued that near the phase transition more fundamental building blocks, namely elementary excitations known as spinons, 
constitute both order parameters. In this scenario staggered Berry phases interfere destructively with the hedgehogs (magnetic monopoles 
in spin space), leading to spinon deconfinement at the phase transition \cite{Senthil-2004}. This mechanism has been confirmed by large 
scale Monte Carlo (MC) simulations of an easy-plane antiferromagnet \cite{Kragset}.       

Previous analyses of spinon deconfinement considered a model for easy-plane antiferromagnets exhibiting a  $U(1)$ symmetry\cite{Motrunich}. The 
resulting theory is described by a  CP$^1$ model, which due to the easy-plane anisotropy has a global $U(1)$ symmetry in addition to the local 
one \cite{Senthil-2004,Motrunich}. This property makes the model self-dual, and it was argued that this self-duality would imply a second-order 
quantum phase transition at zero temperature \cite{Senthil-2004}. However,  Monte Carlo (MC) simulations \cite{Kragset,Kuklov_2006} revealed a 
first-order phase transition, also found by a subsequent renormalization group (RG) analysis of the model \cite{Nogueira_2007}.  Thus, one may 
ask if the same would also be found in the globally $SU(2)$ invariant case supposed to be described by an isotropic CP$^1$ model with a non-compact 
Maxwell term. MC simulations on such a model on a lattice have found a first-order phase transition \cite{Kuklov_2008}. This contradicts previous 
MC results \cite{Motrunich_2008} obtaining a second-order phase transition. Numerical work on the $J-Q$ model appears to show a second-order phase 
transition \cite{Sandvik_2007,Melko-Kaul_2008,Sandvik-2010,Damle}, see however Ref. \onlinecite{Jiang_2008}. 

Recently \cite{Sandvik-2010,Damle}, a feature of the $J-Q$ model which does not seem to follow from DQC, was observed. 
Namely, logarithmic violations of scaling in the zero-temperature spin stiffness and in the finite-temperature
uniform susceptibility, were found. In systems with continuous symmetries, this normally occurs 
either at the upper or lower critical dimensions. In Refs. \cite{Sandvik-2010} and 
\cite{Damle}, the systems considered are $2+1$-dimensional. Thus, the corresponding field theory prescribed by DQC  
would be neither at the upper nor the lower critical dimension. Moreover, no logarithms are expected to occur in the 
zero-temperature spin stiffness or in the finite-temperature uniform susceptibility, since both these quantities can be 
derived from a current correlation function. For this reason, it has been suggested \cite{Sandvik-2010} that the 
DQC scenario should be revised in order to accommodate this new aspect. A phenomenological theory at finite temperature 
involving a gas of free spinons has been proposed recently \cite{Kotov} to fit the logarithmic behavior of the simulations. 
A non-standard power behavior for the thermal gap at criticality was introduced to make a logarithm appear in the 
uniform susceptibility. Since the free spinon gas has the usual spectrum at zero temperature, no quantum critical 
logarithmic behavior can be derived in this way for the spin stiffness at zero temperature. Moreover, the origin of 
the anomalous scaling of the thermal gap was not addressed. In this paper, we show that the logarithmic violation of 
scaling in the $J-Q$ model is actually encoded in the DQC scenario and that this is a direct consequence of 
the emergent $U(1)$ gauge symmetry. 

\begin{figure}
\begin{center}
\includegraphics[width=7cm]{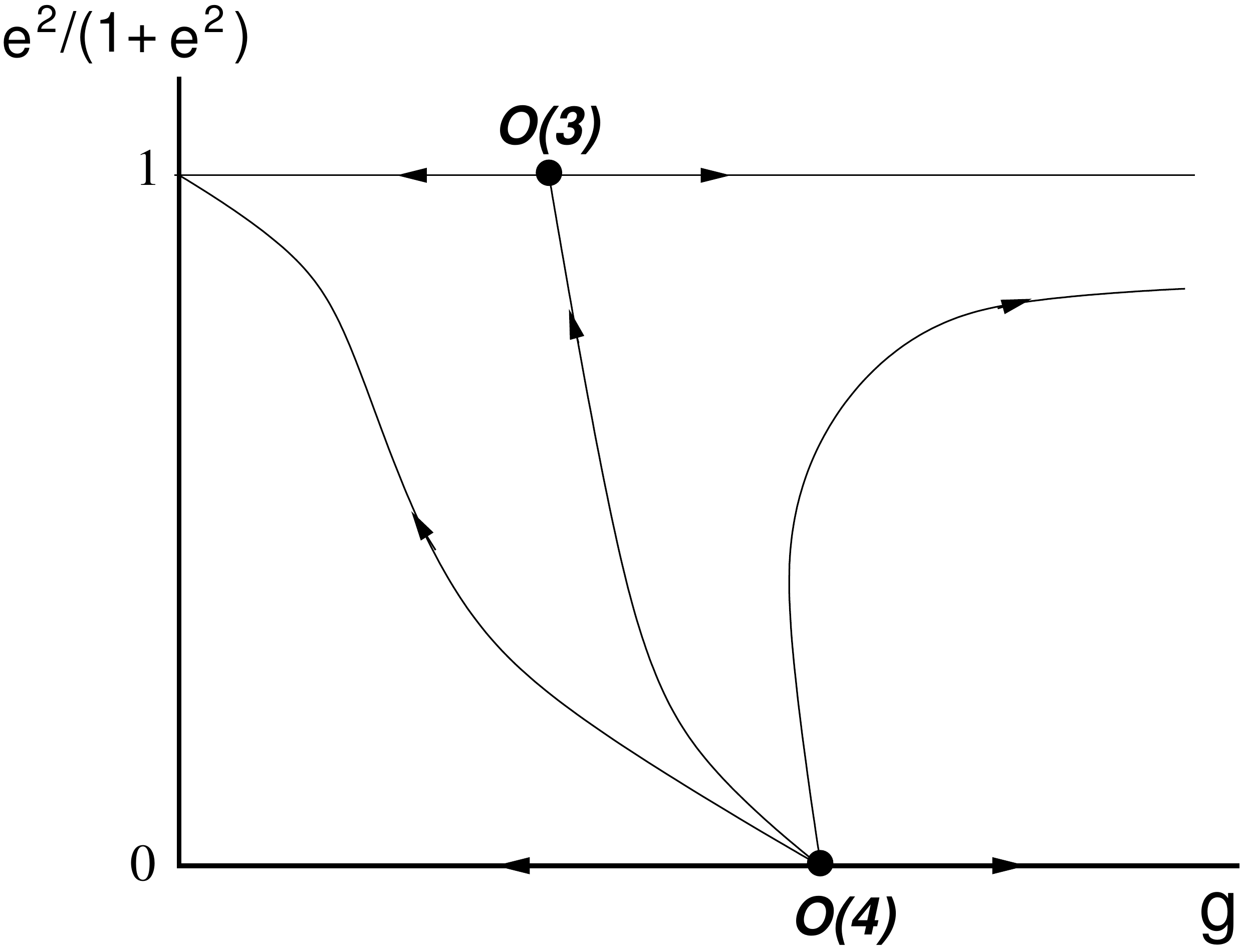}
\caption{Schematic flow diagram of the $CP^1$ model with a compact Maxwell term. The Maxwell term 
allows for an interpolation between the $O(3)$ and $O(4)$ nonlinear $\sigma$ model fixed points.}
\label{Fig:cp1-flow} 
\end{center} 
\end{figure}

If quantum criticality in the $J-Q$ model follows from DQC, it should be governed by an effective lattice 
gauge theory where a staggered Berry phase suppresses its magnetic monopoles \cite{Senthil-2004}. In 
the absence of this staggered Berry phase this lattice gauge theory is given by  
the CP$^1$ model with a compact Maxwell term \cite{Senthil-2004}, 
\begin{equation}
\label{ccp1-lat}
 S=-\frac{1}{g}\sum_{j,\mu,a}z_{aj}^*e^{-iA_{j\mu}}z_{aj+\hat \mu}+{\rm h.c.}-\frac{1}{e^2}\sum_{j,\mu,\nu,\lambda}\cos(\epsilon_{\mu\nu\lambda}
\Delta_{\nu}A_{j\lambda}),
\end{equation}
where $a=1,2$, $\Delta_\mu$ denotes the $\mu$-th component of the lattice gradient, 
and the complex scalar fields satisfy the local constraint $|z_{1,j}|^2+|z_{2,j}|^2=1$. The 
action (\ref{ccp1-lat}) has been studied numerically \cite{Matsui_2005}, and a second order phase transition 
was found. Also, the field theory of Eq. (\ref{ccp1-lat}) has been studied via a renormalization group analysis \cite{Lawrie_1983}. 
The resulting flow diagram interpolates between the quantum critical points of the $O(3)$ and $O(4)$ non-linear 
$\sigma$ models, see Fig. \ref{Fig:cp1-flow}. We thus expect the universality class of the phase transition 
of Eq. (\ref{ccp1-lat}) to be $O(3)$, which obviously features a quantum critical point, in agreement 
with Ref. \cite{Matsui_2005}. 

Within the DQC paradigm, the effect of the staggered Berry phase is essential. This can conveniently be accounted for  
by rewriting the Maxwell term in Eq. (\ref{ccp1-lat}) in Villain form, 
%
 $ S_{\rm Maxwell}=\frac{1}{2e^2}\sum_j(\epsilon_{\mu\nu\lambda}
\Delta_{\nu}A_{j\lambda}-2\pi n_{j\mu})^2$,
and coupling the integer fields $n_{j\mu}$ to fixed, time-independent fields $\zeta_j$ taking the values 0, 1, 2, 3 on 
the dual lattice, as follows $i(\pi/2)\sum_j\zeta_j\Delta_\mu n_{j\mu}$. This defines the Berry 
phase of a compact CP$^1$ model believed to describe the essential  physics at deconfined quantum critical 
points \cite{Sachdev-Jalabert}. A partial dualization of the model  yields \cite{Sachdev-Jalabert}
\begin{eqnarray}
 S_{\rm SJ}&=&\frac{1}{4}\sum_j\left[e^2\left(N_{j\mu}-\frac{1}{4}\Delta_\mu\zeta_j\right)^2
+g(\epsilon_{\mu\nu\lambda}\Delta_\nu N_{j\lambda})^2\right]
\nonumber\\
&+&S_{\bf n}-2\pi i\sum_j N_{j\mu}k_{j\mu}({\bf n}).
\label{SJ_model}
\end{eqnarray}
Here, $S_{\bf n}$ is the action of the nonlinear $\sigma$ model and $k_{j\mu}$ is the topological current. 
In the continuum limit, we have
\begin{equation}
 k_\mu\approx\frac{1}{4\pi}\epsilon_{\mu\nu\lambda}{\bf n}\cdot(\partial_\nu{\bf n}\times\partial_\lambda{\bf n}).
\end{equation}
The lattice 
fields $N_{j\mu}$ are integer-valued and $\zeta_j$ are fixed fields arising from the Berry phase in 
the original model having specific values on the dual lattice (details can be found in Refs. \cite{Sachdev-Jalabert} and 
\cite{RS}). Using the Poisson formula to promote the integer fields $N_{j\mu}$ to real fields $B_{j\mu}$, 
performing the shift $B_{j\mu}\to B_{j\mu}+\Delta_\mu\zeta_j/4$, and integrating over $B_{j\mu}$, one obtains 
\begin{eqnarray}
\label{SJ-dual}
 \tilde S_{\rm SJ}&=&2\pi^2\sum_{i,j}D_{\mu\nu}(x_i-x_j)[k_{i\mu}({\bf n})+m_{i\mu}][k_{j\nu}({\bf n})+m_{j\nu}]
\nonumber\\
&+&S_{\bf n} -\frac{i\pi}{2}\sum_jm_{j\mu}\Delta_\mu\zeta_j,
\end{eqnarray}
Here, $D_{\mu\nu}(x_i-x_j)$  satisfies the equation 
$[(-g\Delta^2+e^2)\delta_{\mu\lambda}+2\Delta_\mu\Delta_\lambda]D_{\lambda\nu}(x_i-x_j)=2\delta_{\mu\nu}\delta_{ij}$,
$m_{j\mu}$ are new integer-valued vector fields arising from the Poisson summation, and we have used 
$\Delta_\mu k_{j\mu}({\bf n})=0$. The first term in Eq. (\ref{SJ-dual}) can be approximated by
\begin{equation}
\label{LM}
 {\cal L}_{\rm Skyrme}=\frac{1}{2e^2}\left[\epsilon_{\mu\nu\lambda}{\bf n}
\cdot(\partial_\nu{\bf n}\times\partial_\lambda{\bf n})\right]^2,
\end{equation}
first introduced in Ref. \onlinecite{Skyrme}.  The effective Lagrangian in the continuum limit is thus
\begin{equation}
\label{FS}
 {\cal L}=\frac{1}{2g}(\partial_\mu{\bf n})^2+{\cal L}_{\rm Skyrme}+\dots,
\end{equation}
where the three-component direction field ${\bf n}$ satisfies the local constraint ${\bf n}^2=1$ and the ellipses denote 
other terms related to the Berry phase, which are being neglected in the above approximation. We return to this below. The 
model in Eq. (\ref{FS}) \cite{Faddeev_1970} is known to have a rich geometric and topological structure 
\cite{Faddeev-Niemi_Nature,Babaev-PRB_2002,Babaev-PRB_2009}. The emergent $U(1)$ symmetry of the Skyrme term follows from 
the compact $U(1)$ gauge group in the continuum arising as a subgroup of the $SU(2)$ gauge group \cite{Polyakov}. To see 
this, consider the functional integral
\begin{equation}
 Z=\int{\cal D}{\bf n}{\cal D}C_\mu \exp\left(-\frac{1}{4e^2}\int d^3x{\bf F}_{\mu\nu}\cdot{\bf F}_{\mu\nu}\right),
 \label{func_int}
\end{equation}
where ${\bf F}_{\mu\nu}=\partial_\mu{\bf W}_\nu-\partial_\nu{\bf W}_\mu-{\bf W}_\mu\times{\bf W}_\nu$ is an $SU(2)$ field 
strength with an adjoint non-Abelian gauge field of the form 
${\bf W}_\mu={\bf n}\times\partial_\mu{\bf n}+{\bf n}C_\mu$, and $C_\mu$ is an Abelian gauge field. The Skyrme contribution  
Eq. (\ref{LM}) follows from integrating out $C_\mu$ in Eq. \ref{func_int}. This is reminiscent of the arguments of Ref. 
\onlinecite{Faddeev-Niemi_PRL_1999}, where a four-dimensional version of Eq. (\ref{FS}) \cite{Note-FS} is 
argued to be a low-energy description of $SU(2)$ Yang-Mills theory.    

The model Eq. (\ref{FS}) has to be modified to account for the destructive interference between the Berry phases and the 
magnetic monopoles. As discussed in detail in Ref. \cite{Senthil-2004}, this interference mechanism suppresses the monopoles, 
which implies an effective model given by the {\em non-compact} CP$^1$ model with a Maxwell term. Such a model may be written 
on the form \cite{Babaev-PRB_2009} ,
\begin{equation}
\label{FSB}
 {\cal L}=\frac{1}{2g}(\partial_\mu{\bf n})^2+\frac{1}{2g}C_\mu^2+\frac{1}{2e^2}[\epsilon_{\mu\nu\lambda}\partial_\nu C_\lambda
+\epsilon_{\mu\nu\lambda}{\bf n}
\cdot(\partial_\nu{\bf n}\times\partial_\lambda{\bf n})]^2.
\end{equation}
Setting ${\bf n}=z_a^*\sigmab_{ab}z_b$, we obtain the CP$^1$ realization of Eq. \ref{FSB}, 
\begin{equation}
\label{CP1}
 {\cal L}=\frac{1}{g}|(\partial_\mu-iA_\mu)z_a|^2+\frac{1}{2g}C_\mu^2+\frac{1}{2e^2}
(\epsilon_{\mu\nu\lambda}\partial_\nu C_\lambda+2i\epsilon_{\mu\nu\lambda}\partial_\nu z_a^*\partial_\lambda z_a)^2.
\end{equation}
Classically, by enforcing the constraint $|z_1|^2+|z_2|^2=1$, we have, 
\begin{equation}
\label{A}
 A_\mu=(i/2)(z_a^*\partial_\mu z_a-z_a\partial_\mu z_a^*).
\end{equation}
We can now perform the {\it singular} gauge transformation  
$A_\mu\to A_\mu - C_\mu$, $z_a\to e^{i\int_0^x dx_\mu^{'}C_\mu(x')}z_a$,  to obtain,
\begin{equation}
 \label{CP1-1}
 {\cal L}=\frac{1}{g}|(\partial_\mu-iA_\mu)z_a|^2+\frac{1}{2e^2}
(\epsilon_{\mu\nu\lambda}\partial_\nu A_\lambda)^2.
\end{equation}
This is precisely the standard DQC model \cite{Senthil-2004}. Note that a similar decoupling  does not hold in the case of 
the Ginzburg-Landau theory for two-component superconductors discussed in Ref. \cite{Babaev-PRB_2009}, since there the sum 
of the respective Cooper pair densities is not $CP^1$-constrained. 
Thus, we see that if we insist on the emergent character of $A_\mu$ as expressed in Eq. (\ref{A}), the Maxwell 
term in Eq. (\ref{CP1-1}) is just the Skyrme term, which this is shown to be contained in Eq. (\ref{SJ_model}).  

The gauge transformation employed to decouple $C_\mu$ and thus derive Eq. (\ref{CP1-1}) was performed at the classical 
level, in which case the equation of motion for $A_\mu$ yields $A_\mu=(i/2)(z_a^*\partial_\mu z_a-z_a\partial_\mu z_a^*)$. 
In some calculations involving higher order quantum fluctuations, it might be more appropriate to consider the 
Lagrangian (\ref{CP1}), since quantum fluctuations give the gauge field $A_\mu$ an independent dynamics. However, 
in most lowest order approximations, Eq. (\ref{CP1-1}) is sufficiently accurate. Note that Eq. (\ref{A}) does not 
follow from the equation of motion for $A_\mu$ derived from the Lagrangian (\ref{CP1-1}). 
  
Next we calculate the spin stiffness, $\rho_s$. The latter is obtained from the gauge invariant response to a 
twist associated to the spin current tensor ${\bf J}_\mu=({\bf n}\times\partial_\mu{\bf n})/g$ \cite{Polyakov-Book}. Due to the constraint 
${\bf n}^2=1$, we have, 
\begin{equation}
\frac{g}{2}{\bf J}_\mu^2=\frac{1}{2g}(\partial_\mu{\bf n})^2
\end{equation}
which provides some insight into the meaning of $1/g$ as the {\it bare} stiffness. Note that when $d=1+0$, ${\bf J}_\mu$ is just 
the angular momentum of a particle having moment of inertia $1/g$ and constrained to move on the surface of the $S^2$ 
sphere. By generalizing the mechanics of a particle on a sphere to a field theory in 
$d=D+1$ spacetime dimensions, the twist is realized by the response to an external constant triplet field ${\bf S}_\mu$, which amounts to 
makiing the replacement $\partial_\mu{\bf n}\to \partial_\mu{\bf n}+{\bf S}_\mu\times{\bf n}$.  
In order to facilitate the calculations, we can assume without loss of generality that $S_\mu^a=\delta_{a=3}S_\mu$. 
Thus, in the CP$^1$ representation $S_\mu$ will couple to  the third component of ${\bf J}_\mu$, 
which is written in terms of spinon fields as 
%
 $ J_\mu^3=\frac{1}{g}[j_\mu^{(1)}-j_\mu^{(2)}]$,
where 
%
 $ j_\mu^{(a)}=i(z_a^*\partial_\mu z_a-z_a\partial_\mu z_a^*)-2A_\mu|z_a|^2$,
with no sum over $a$. Variation of the Lagrangian Eq. (\ref{CP1-1}) yields,
%
 $\frac{1}{g}[j_\mu^{(1)}+j_\mu^{(2)}]=\frac{1}{e^2}\partial_\nu F_{\nu\mu}$, 
which allows us to write
%
 $J_\mu^3=\frac{2}{g}j_\mu^{(1)}-\frac{1}{e^2}\partial_\nu F_{\nu\mu}$.
The spin stiffness is calculated by computing the response of $J_\mu^3$ when it is coupled to an external source, $S_\mu$. 
Differentiating the free energy functional with respect to the source, letting the source vanish at the end, yields the spin 
stiffness in the form, 
\begin{equation}
 \rho_s=\frac{4}{g}\langle|z_1|^2|z_2|^2\rangle-\frac{1}{d}\int d^dxK_{\mu\mu}(x),
\end{equation}
where $K_{\mu\nu}(x)=\langle J_\mu^3(x)J_\nu^3(0)\rangle$ is the current correlator. The above formula is a generalization to 
the $O(3)$ case of the well-known formula for the superfluid density of a globally $U(1)$-invariant system. Interestingly, the 
$O(3)$ stiffness involves the correlator of a gauge-invariant $U(1)$ current within a framework associated to a global $SU(2)$ symmetry.   

When only global symmetries are involved, conserved currents do not renormalize \cite{Gross-1976,Collins}. This fact
elegantly provides a foundation for the scaling relation $\rho_s \sim \xi^{2-D}$ for the superfluid stiffness in 
systems with a global $U(1)$ symmetry \cite{Josephson}. Indeed, current conservation implies 
that the current correlation function does not exhibit an anomalous dimension, whence the scaling behavior of 
the superfluid stiffness is simply determined by dimensional analysis. However, the non-renormalization theorem 
fails in gauge theories \cite{Collins-1}, such as the Lagrangian in Eq. (\ref{CP1-1}). In the absence of the Maxwell, 
we have the standard CP$^1$ model, which is equivalent to the $O(3)$ nonlinear sigma model. In this case the non-renormalization 
theorem is still valid, since the gauge field then is just an auxiliary field. Specifically, if we introduce the dimensionless 
coupling $\hat g=g\Lambda^{d-2}$, where $\Lambda$ is the ultraviolet cutoff, along with the parameter $r=1-\hat g/\hat g_c$, with 
$\hat g_c$ being the critical coupling, we obtain that for the standard CP$^1$ model in $d=D+1$ spacetime dimensions 
near criticality, $\rho_s\sim r^{\nu(d-2)}$, corresponding to 
standard Josephson scaling \cite{Josephson}. This result can be derived using an expansion in $\epsilon=d-2$ or, in the case of 
the CP$^{N-1}$ model, by means of an $1/N$ expansion \cite{Kaul-Sachdev}. Furthermore, at finite temperature and at $\hat g=\hat g_c$, scale 
invariance implies $\rho_s\sim T^{d-2}$, which is essentially the quantum critical spin susceptibility at finite temperature.     

For the model in Eq. (\ref{CP1-1}), the 
presence of the Maxwell term leads to logarithmic corrections in the spin stiffness for $d=2+1$ when 
approaching the critical point from the broken symmetry (Higgs) phase, where 
the gauge field is gapped. More precisely, the exact expression up to order $1/N$ and $d=2+1$, 
keeping both $Ng$ and $Ne^2$ fixed is given by \cite{Nogueira-Sudbo-2012},
\begin{widetext}
\begin{equation}
\label{rhos-full}
 \frac{\rho_s}{r}   \sim   1-\frac{16}{3\pi^2N}\ln\left(\frac{16r}{N\hat g+16r}\right)
+\frac{64}{3\pi^2N\left(1+\frac{256r}{N^2\hat f\hat g}\right)}\left[\ln\left(\frac{64r}{N^2\hat f\hat g}\right)
+\frac{3}{\sqrt{1-\frac{2048r}{N^2\hat f\hat g}}}\ln\left(\frac{1+\sqrt{1-\frac{2048r}{N^2\hat f\hat g}}}
{1-\sqrt{1-\frac{2048r}{N^2\hat f\hat g}}}\right)\right]+\dots,
\end{equation}
\end{widetext}
where we have introduced the dimensionless coupling $\hat f=e^2/\Lambda$. For $\hat f\gg 1$, we obtain, 
\begin{eqnarray}
 \frac{\rho_s}{r}  
\approx  
1 - \frac{16}{3\pi^2N}\ln\left(\frac{16r}{N\hat g+16r}\right)
  - \frac{128}{3\pi^2N} \ln\left(\frac{128r}{N^2\hat f\hat g}\right),
\end{eqnarray}
which implies the critical exponent $\nu=1-48/(\pi^2N)$ of the standard CP$^{N-1}$ model at large $N$ \cite{Kaul-Sachdev}. 
The same behavior is obtained at fixed $\hat f$ and $r\to 0$, consistent with the result that at the CP$^{N-1}$ fixed point 
the Maxwell term is (dangerously) irrelevant.  
For $\hat f\to 0$, on the other hand, only the first two terms in Eq. (\ref{rhos-full}) remain, leading 
to the critical exponent $\nu=1-16/(3\pi^2 N)$ of an $O(2N)$ non-linear sigma model. Therefore, Eq. (\ref{rhos-full}) 
interpolates between the fixed points of the CP$^{N-1}$ and $O(2N)$ models. The logarithmic correction to the Josephson 
scaling obtained here is in agreement with recent numerical findings \cite{Sandvik-2010}. This provides further evidence 
that the DQC scenario describes the quantum critical regime of the AF-VBS transition in low-dimensional quantum spin 
models. These log-corrections, appearing away from lower and upper critical dimensions, reflect an emergent gauge symmetry 
of such systems. They also elucidate the dangerously irrelevant character of the emergent Maxwell term in the CP$^{N-1}$ 
model, and its primary role in the log-correction.   

The interpolation between the large $N$ fixed points of the CP$^{N-1}$ and $O(2N)$ obtained above is of 
the same type we have found before in our discussion in connection with Fig. \ref{Fig:cp1-flow}. For
$N=2$ the standard CP$^{N-1}$ model is equivalent to the $O(3)$ non-linear sigma model. The DQC regime lies on  
the critical separatrix of Fig. \ref{Fig:cp1-flow}. Thus, the logarithmic violation of scaling obtained 
here is fundamentally different from the one usually encountered at the upper or lower critical dimension of local 
field theories. 

It would be interesting to observe the violation of Josephson scaling in experiments by measuring the spin susceptibility, which 
would essentially be a measurement of the spin stiffness. Good candidates are lowdimensional quantum antiferromagnets featuring 
geometric frustration, such as the organic Mott insulating compound EtMe$_3$P[Pd(dmit)$_2$]$_2$ \cite{Kato-2007} or the Kagom\'e 
lattice system Zn$_x$Cu$_{4-x}$(OH)$_6$Cl$_2$ \cite{Lee-NatMat}. Field theories with emergent $U(1)$ gauge symmetries have been 
proposed to describe the quantum criticality in these materials  \cite{Xu-2010,Lawler-PRL-2008}.         

In this paper, we have considered a class of quantum spin systems featuring a breakup of spin-1/2 objects into more 
fundamental constituents called spinons, accompanied by an emergent massless gauge-field and hence an emergent $U(1)$ gauge
symmetry. We show that this emergence gives rise to logarithmic violations of Josephson scaling of the spin stiffness of these
systems. It originates with a breakdown of the non-renormalization property of the conserved current that underpins Josephson scaling. 
Similar violations of scaling should appear in the susceptibility of the system. A violation of Josephson scaling has been observed 
in numerical works on the spin stiffness of quantum antiferromagnets with ring-exchange, and constitutes an experimental signature of 
spin-fractionalization and emergence of massless photons in low-dimensional quantum antiferromagnets. We have proposed candidate 
materials in which to look for this.  

\acknowledgments

The authors thank V. Kotov, O. I. Motrunich, A. Sandvik, U.-J. Wiese, and E. Babaev, for discussions. F.S.N. acknowledges support of 
the Deutsche Forschungsgemeinschaft (DFG) through the Collaborative Research Center SFB Transregio 12. A. S. acknowledges support 
from the Norwegian Research Council Grants No. 205591/V30 (FRINAT).


\begin{thebibliography}{100}

\bibitem{Coleman} S. Coleman, {\it Aspects of Symmetry} (Cambridge University Press, 1985); P. W. Anderson, 
{\it Basic Notions of Condensed Matter Physics} (Westview Press, 1997).

\bibitem{Savit} R. Savit, Rev. Mod. Phys. {\bf 52}, 453 (1980).

\bibitem{Kleinert-GFCM-1} H. Kleinert, {\it Gauge Fields in Condensed
Matter}, Vols. 1 and 2 (World Scientific, Singapore, 1989).

\bibitem{Sudbo-Book} K. Fossheim and A. Sudb\o, {\it Superconductivity: Physics and Applications} 
(John Wiley \& Sons, 2004).


\bibitem{Sachdev-Review} S. Sachdev, Nature Physics {\bf 4}, 173 (2008) and references therein.

\bibitem{Sandvik_2007} A. W. Sandvik, Phys. Rev. Lett. {\bf 98}, 227202 (2007).

\bibitem{Melko-Kaul_2008} R. G. Melko and R. K. Kaul, Phys. Rev. Lett. {\bf 100}, 017203 (2008).  

\bibitem{Jiang_2008} F.-J. Jiang, M. Nyfeler, S. Chandrasekharan, and U.-J. Wiese, 
J. Stat. Mech. P02009 (2008).

\bibitem{Sandvik-2010} A. W. Sandvik, Phys. Rev. Lett. {\bf 104}, 177201 (2010).

\bibitem{Senthil-2004} T. Senthil, A. Vishwanath, L. Balents, S. Sachdev,
 and M. P. A. Fisher, Science, {\bf 303}, 1490 (2004); T. Senthil, L. Balents, S. Sachdev,
A. Vishwanath, and M. P. A. Fisher, Phys. Rev. B {\bf 70}, 144407 (2004).

\bibitem{Kragset} S. Kragset, E. Sm{\o}rgrav, J. Hove, F. S. Nogueira, and A. Sudb{\o}, 
Phys. Rev. Lett. {\bf 97}, 247201 (2006).

\bibitem{Motrunich} O. I. Motrunich and A. Vishwanath, Phys. Rev. B  {\bf 70}, 
075104 (2004).

\bibitem{Kuklov_2006} A. B. Kuklov, N. V.  Prokof'ev, B. V. Svistunov,
and M. Troyer, Ann. Phys. (N.Y.)  {\bf 321}, 1602 (2006).

\bibitem{Nogueira_2007} F. S. Nogueira, S. Kragset, and A. Sudb{\o}, Phys. Rev. B {\bf 76}, 220403(R) (2007). 

\bibitem{Kuklov_2008} A. B. Kuklov, M. Matsumoto, N. V.  Prokof'ev, B. V. Svistunov,
and M. Troyer,  Phys. Rev. Lett. {\bf 101}, 050405 (2008). 

\bibitem{Motrunich_2008} O. I. Motrunich and A. Vishwanath, arXiv:0805.1494; see also 
A. B. Kuklov, M. Matsumoto, N. V.  Prokof'ev, B. V. Svistunov,
and M. Troyer, arXiv:0805.2578. 

\bibitem{Damle} A. Banerjee, K. Damle, and F. Alet, 
Phys. Rev. B {\bf 82}, 155139 (2010).  

\bibitem{Kotov}  A. W. Sandvik, V. N. Kotov, and O. P. Sushkov, Phys. Rev. Lett. {\bf 106}, 207203 (2011).  






\bibitem{Matsui_2005} S. Takashima, I. Ichinose, and T. Mastui, Phys. Rev. B {\bf 72}, 075112 (2005).

\bibitem{Lawrie_1983} I. D. Lawrie and C. Athorne, J. Phys. A: Math. Gen. {\bf 16}, 
L587 (1983); {\it Ibid.} {\bf 16}, 4428 (1983).

\bibitem{Sachdev-Jalabert} S. Sachdev and R. Jalabert, Mod. Phys. Lett. B {\bf 4}, 1043 (1990).

\bibitem{RS} N. Read and S. Sachdev, Phys. Rev. Lett. {\bf 62}, 
1694 (1989); Phys. Rev. B {\bf 42}, 4568 (1990).

\bibitem{Skyrme} T. H. R. Skyrme, Proc. R. Soc. Lond. A {\bf 260}, 127 (1961).

\bibitem{Faddeev_1970} L. D. Faddeev, Preprint IAS-75-QS70 (Institute for Advanced Studies, Princeton, 1970).

\bibitem{Faddeev-Niemi_Nature} L. D. Faddeev and A. J. Niemi, Nature (London) {\bf 387}, 58 (1997).

\bibitem{Babaev-PRB_2002} E. Babaev, L. D. Faddeev, and A. J. Niemi, Phys. Rev. B {\bf 65}, 100512(R) (2002).

\bibitem{Babaev-PRB_2009} E. Babaev, Phys. Rev. B {\bf 79}, 104506 (2009).

\bibitem{Polyakov} A. M. Polyakov, Nucl. Phys. B {\bf 120}, 429
(1977).

\bibitem{Faddeev-Niemi_PRL_1999} L. D. Faddeev and A. J. Niemi, Phys. Rev. Lett. {\bf 82}, 1624 (1999).

\bibitem{Note-FS} In order to generalize (\ref{LM}) to 3+1 dimensions, just note that Eq. (\ref{LM}) can 
be equivalently written as 
${\cal L}_{\rm Skyrme}=(1/2e^2)[(\partial_\mu{\bf n})^2(\partial_\nu{\bf n})^2-(\partial_\mu{\bf n}\cdot\partial_\nu{\bf n})^2]$. 










\bibitem{Polyakov-Book} A. M. Polyakov, {\it Gauge Fields and Strings} (Harwood Academic Publishers GmbH, Chur, Switzerland, 1987). 








\bibitem{Gross-1976} D. J. Gross, ``Applications of the renormalization group'' in: {\it Methods in Field Theory}, R. Balian and 
J. Zinn-Justin (Eds.) (North-Holland, Amsterdam, 1976).  

\bibitem{Collins} J. C. Collins, {\it Renormalization} (Cambridge University Press, Cambridge, 1984).

\bibitem{Josephson} B. D. Josephson, Phys. Lett. {\bf 21}, 608 (1966).

\bibitem{Collins-1} J. C. Collins, A. V. Manohar, and M. B. Wise, Phys. Rev. D {\bf 73}, 105019 (2006).  

\bibitem{Kaul-Sachdev} R. K. Kaul and S. Sachdev,  Phys. Rev. B {\bf 77}, 155105 (2008).

\bibitem{Nogueira-Sudbo-2012} F. S. Nogueira and A. Sudb{\o} (in preparation). 


\bibitem{Kato-2007} Y. Shimizu, H. Akimoto, H. Tsujii, A. Tajima, and R. Kato, Phys. Rev. Lett. {\bf 99}, 256403 (2007).

\bibitem{Lee-NatMat} S.-H. Lee et al., Nat. Mater. {\bf 6}, 853 (2007). 

\bibitem{Xu-2010} C. Xu and S. Sachdev, Phys. Rev. B {\bf 79}, 064405 (2009).

\bibitem{Lawler-PRL-2008} M. J. Lawler, L. Fritz, Y. B. Kim1, and S. Sachdev, Phys. Rev. Lett. {\bf 100}, 187201 (2008).

\end{thebibliography}
\end{document}